\documentstyle[prd,aps,epsfig,floats]{revtex}
\begin{document}

\draft
\input epsf \renewcommand{\topfraction}{0.8}
\renewcommand{\topfraction}{0.8} 
\newcommand{\be}{\begin{equation}}
\newcommand{\ee}{\end{equation}}
\newcommand{\bea}{\begin{eqnarray}}
\newcommand{\beas}{\begin{eqnarray*}}
\newcommand{\eea}{\end{eqnarray}}
\newcommand{\eeas}{\end{eqnarray*}} 
\newcommand{\ba}{\begin{array}}
\newcommand{\ea}{\end{array}}
\newcommand{\pbar}{\not{\!\partial}}
\newcommand{\dbar}{\not{\!{\!D}}}
\def\lsim{\:\raisebox{-0.75ex}{$\stackrel{\textstyle<}{\sim}$}\:}
\def\gsim{\:\raisebox{-0.75ex}{$\stackrel{\textstyle>}{\sim}$}\:}

\title{Scenarios of modulated perturbations}

\author{Rouzbeh Allahverdi}
\address{Theory Group, TRIUMF, 4004 Wesbrook Mall, Vancouver, B.C., V6T 
2A3, Canada.}
\maketitle
\begin{abstract}

In an alternative mechanism recently proposed, adiabatic cosmological
perturbations are generated at the decay of the inflaton field due to
small fluctuations of its coupling to matter. This happens whenever
the coupling is governed by the vacuum expectation value of another
field, which acquires fluctuations during inflation. We
discuss generalization and various possible implementations of this 
mechansim, and present some specific particle physics examples. In many cases 
the 
second field can start oscillating before perturbations are imprinted, or 
survive long enough so to dominate over the decay products of the inflaton. 
The primordial perturbations will then be modified accordingly in each case. 

\end{abstract}

\vskip2pc

\maketitle


\section{Introduction}

The detection of the cosmic microwave
background (CMB) anisotropies indicates the presence of coherence over
super horizon scales, which is a strong indication of an early
inflationary stage~\cite{inflation}. To date measurements are in agreement 
with the
simplest prediction from inflation which is a nearly scale invariant
spectrum of gaussian and adiabatic primordial
perturbations~\cite{wmap}. Nevertheless, possible experimental or theoretical
deviations from this most economical possibility have been the subject
of intense study.  Indeed, the increasing precision of the CMB
measurements gives us the hope to raise the bar in the near future from
the present confirmation of the general idea of inflation, to the
study and determination of the underlying theory.

An important aspect of inflation is the stage of reheating~\cite{reheat}, which
describes all the processes from the end of the inflationary expansion
to the following hot ``big-bang'' evolution. The inflaton decay 
is only the first, although most relevant, stage of this process. The
inflaton decay is typically very quick and non-perturbative, and it
occurs immediately after the end of inflation~\cite{preheat}. Due to its 
efficiency,
it leads to a distribution of particles which is very far from thermal
equilibrium, so that reheating completes over a much longer
timescale. Particle physics plays a relevant role in this period. It
seems conceivable that the baryon asymmetry of the universe is
generated at this stage. On the other hand, nucleosynthesis poses
strong bounds on the production of gravitationally decaying relics. These
requirements can be combined to constrain different models of
reheating. For example, limits from nucleosynthesis are in
contrast with thermal grand unified theory (GUT) baryogenesis and only 
marginally compatible
with thermal leptogenesis, while a non-thermal origin~\cite{non} can be more
easily accounted for. However, while these considerations allow us to
study the inflaton interactions for any given particle physics model,
we are not yet in a stage where we can discriminate among different
scenarios. As in many other areas of physics, we are more in the need
of experimental evidence rather than theoretical models. It is
therefore important to ask whether reheating can have some other
observational consequences which can guide us discriminating among the
different possibilities.

A positive answer to this question is provided by the recent
observation that reheating could have played a key role also for generating the
primordial perturbations leaving their imprints on the CMB, as well as on the
matter power spectrum~\cite{dgz,kofman}. This statement challenges the
common assumption that the microphysics responsible for the decay of
the inflaton, and the successive thermalization, cannot
have an impact on the much larger scales relevant for the present
observations which at that time are well outside the horizon. However,
perturbations generated during inflation
can be strongly affected by the following history of the universe. For
example, the adiabatic mode of the perturbations is sensitive to the
evolution of dominating species, and it is precisely during
reheating that this evolution is mostly unknown. Although we can be
sure of the starting (unstable inflaton) and final (thermal bath)
points, we do not know what dominates during the intermediate stages
and how its equation of state evolves. Several possibilities can
occur, depending on the inflaton potential near its minimum (for
example, $m^2 \phi^2$ versus $\lambda \phi^4$ potential), on
the possible generation of unstable heavy particles which can dominate for
some time, on the possible phase transitions and on the presence of
large effective masses before thermalization completes.

These processes can strongly affect primordial perturbations if they
occur at (slightly) different moments in the different parts of the
universe we presently observe. This can happen, for example, for the
decay of the inflaton, if its coupling to matter is controlled by the
vacuum expectation value (VEV) of some scalar field $X\,$ which acquires
inflationary fluctuations~\cite{dgz,kofman}. The equation of
state of the inflaton is in general different from the one of its
decay products, and fluctuations in the decay rate will then give rise
to fluctuations in the energy density and in the metric. Perturbations
generated by this mechanism have been named {\em modulated
perturbations}. However, modulated perturbations do not need to be generated
precisely at the inflaton decay. They could also arise due to
fluctuations in the decay rate of some other intermediate particle, or
in the rate of the processes which are more relevant for
thermalization. In general, we will denote by $t_*$ the time at which
this process completes, with the assumption that it can be anytime during
reheating. This process changes the equation of state
of the universe from an unknown one $p=w\rho$ to $p = \rho/3\,$, which is
typical of radiation. Due to the change in $w$,
fluctuations in $t_*$ give rise to adiabatic perturbations. 

A further extension on which we want to focus, noted briefly
in~\cite{kofman}-\cite{gruzinov}, is the
possibility that the field $X$ is evolving before or after
$t_*\,$. While $X$ is initially frozen at some (nearly) constant
value, it will start oscillating around the minimum of its potential (with a 
slowly decreasing amplitude) as soon as the Hubble expansion rate drops 
below its mass. This can significantly affect (typically decrease) the
amplitude of modulated perturbations if oscillations of $X$ start before 
perturbations are generated. Moreover, as any other modulus, $X$ can 
eventually dominate the energy density of the universe. Isocurvature 
fluctuations in $X$ 
will in this case be converted to curvature perturbations in the same way as 
the curvaton 
scenario~\cite{curvaton,moroi}. Then perturbations in the new (dominating) 
thermal bath will replace modulated perturbations. A particular 
identification of $X$, and requirements for a successful reheating, 
may favour one of these cases.           

The analysis of Section~\ref{general} covers all these different
possibilities. From the general discussion, we then specialize to some
specific examples motivated by particle physics in Section~\ref{examples}. 
The main ingredient of the 
modulated perturbations mechansim is the existence of a scalar field with a 
mass less than the Hubble expansion rate during inflation. Theories based on 
supersymmerty provide the best framework for implementing this mechansim as 
they contain many scalar fields whose mass is protected against quantum 
corrections. We first discuss the possibility that $X$ is a modulus
field, with a mass $\sim$ TeV and interactions of gravitational strength
with the observable sector. This is probably the most natural
implementation of the idea of modulated perturbations: moduli of
superstring models have these properties, and their VEV's control
the value of the parameters of the theory. 
The difficulty with this implementation, as we shall
see, is that it typically results in a too small amplitude for the
primordial perturbations, unless $X$ dominates. In this
case, however, one has to make sure that the late (gravitational)
decay of $X$ occurs before primordial nucleosynthesis. Alternative
examples that we consider are the identification of $X$ with
right-handed (RH) sneutrinos and with flat directions of the minimal
supersymmetric standard model (MSSM), where the decay of $X$ occurs at
an earlier stage. In general, other scalar fields (in particular those 
coupled to $X$) should have negligible 
fluctuations in order not to affect perturbations. In Section~\ref{discuss} 
we discuss possible ways to achieve such a suppression leading us to some 
``benchmark scenarios'' of modulated perturbations. The results will 
be summarized in the concluding Section~\ref{end}.


\section{General analysis} \label{general}

In this section we discuss how modulated perturbations are sensitive
to different assumptions on the dynamics of $X\,$.~\footnote{A simliar
analysis when $X$ is the curvaton field has been performed
in~\cite{nicola}, \cite{dim}.} As discussed in the Introduction, we assume 
that the VEV of $X$ controls the rate $\Gamma_* = H \left( t_* \right)$ of
some process at reheating. We can in general expect 
\begin{equation}
\Gamma_* = \Gamma_0 + X^n \, f \left( X \right) \,\,,
\label{rate}
\end{equation}
where $\Gamma_0$ represents any $X$-independent contribution to that process.
Here $n$ is an ${\cal O}\left( 1 \right)$ 
real number, while $f$ is either a (dimensionful) constant or a function 
which depends
only weakly on $X\,$. Eq.~(\ref{rate}) holds for a number of
possibilities. For example, it can be applied when $X$ is included in
the vertex representing some interaction, or when $X$ determines the mass of a
decaying particle. For specific 
examples, consider the following cases of perturbative decay and preheating. 
In a 
perturbative decay, $\Gamma_*$ is a linear function of mass and a quadratic 
function of coupling. Then linear dependence of the coupling and mass on $X$ 
will result in $n=2$ and $n=1$ respectively, and $f$ is a constant. Whilst, 
for 
preheating to bosons,~\footnote{The situation will be different for fermionic 
preheating~\cite{gk},~\cite{peloso}. However, in supersymmetry (which 
provides the 
natural framework for implementing the modulated perturbations mechanism) 
preheating to 
fermions is always accompanied by bosonic preheating which is much more 
efficient.} $\Gamma_*$ is a logarithmic function of coupling and 
linear function of mass~\cite{preheat}. Therefore, linear dependence of mass 
and coupling on $X$ will result in $n=1$ ($f$ being a constant) and $n=0$ 
($f$ a logarithmic function) respectively.
           
Modulated perturbations are generated at $t_*$ with an amplitude
\begin{equation}
\zeta \sim \frac{\delta{\rho}}{\rho} \sim
\frac{\delta t_*}{t_*} \sim {\delta \Gamma_* \over \Gamma_*}
\Big\vert_{t_*} \,\,.
\label{start}
\end{equation}
The overall proportionality factor can be relevant, but 
we prefer to leave it unspecified not to affect the generality of the 
discussion. We assume that the potential for $X$ has a minimum at $X_0\,$, and
that only the first (quadratic) term in an expansion series around
$X_0\,$ is relevant:
\begin{equation}
V \left( X \right) = \frac{1}{2} m_X^2 \left( X - X_0 \right)^2 \,\,,
\label{model}
\end{equation}
where $m_X$ is the mass of $X$. We shall notice that there can be some 
degeneracy between $X_0$ and
$\Gamma_0\,$. For example, consider the case where perturbations are 
generated at the inflaton decay\footnote{Henceforth, we occassionally mention 
the inflaton decay as an explicit example. 
Nevertheless, the main conclusions will hold for 
other processes generating modulated perturbations.} and there are 
two decay modes, one mediated by $X$ and one 
$X-$independent, with the same final state: ${\cal L}
\supset X \phi {\bar \psi} \psi / M + h \phi {\bar \psi} \psi \,$ ($\psi$ 
being a fermion). The
redefinition ${\tilde X} \equiv hM + X$ then sets $\Gamma_0 = 0\,$,
while changing the value of ${\tilde X}_0\,$. 
In what follows, $\Gamma_0$ stands for the $X$-independent part of $\Gamma_*$ 
which cannot 
be removed by a redefinition of $X_0$. This happens, for example, 
when the two decay modes of the inflaton have different final states. 

Let us denote the Hubble rate during inflation by $H_I$. For $m_X >
H_I \,$, the field $X$ settels at the minimum $X_0\,$ with negligible
fluctuations. We then recover the standard situation of a constant
rate $\Gamma = \Gamma \left( X_0 \right) \,$. In the opposite case, $m_X < 
H_I$ and 
fluctuations of $X$ accumulate in a coherent expectation value $X_I$, with a 
dispersion $\delta X \sim H_I \,$. This gives a typical 
result for modulated perturbations~\cite{dgz,kofman}:
\begin{equation}
\zeta \sim H_I / X_I \,\,.
\label{res1}
\end{equation}
\begin{figure}[tb]
\includegraphics{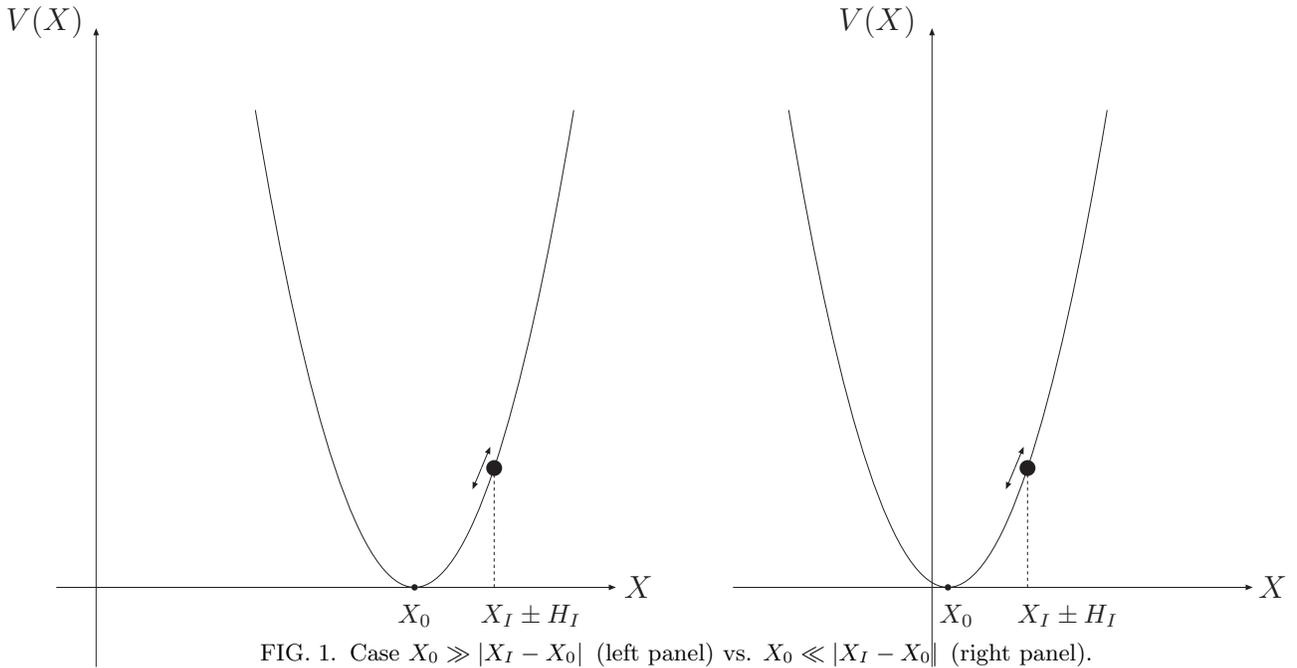}
\caption{Case $X_0 \gg \vert X_I - X_0 \vert\,$ (left panel) vs.  $X_0 \ll
\vert X_I - X_0 \vert\,$ (right panel).}
\label{fig1}
\end{figure}
As discussed
in the Introduction, there are however interesting cases in which,
due to the evolution of $X$ prior or after $t_*\,$, eq.~(\ref{res1})
does not apply. The amplitude of modulated perturbations can be
easily computed case by case and, although there are several
parameters in the model, the final results can be most effectively
summarized as a function of the mass and the decay rate of the
scalar field $X\,$. The two cases $X_0 \gg \vert X_I - X_0
\vert\,$ and $X_0 \ll \vert X_I - X_0 \vert\,$, see fig.~(\ref{fig1}),
lead to different results, and hence we will consider them
separately in the next two subsections.


\subsection{Case I: $\vert X_I - X_0 \vert \ll X_0 \leq M_p$}

Figure~(2) summarizes the results for this case. We always
demand that $\Gamma_X < \Gamma_*$, with $\Gamma_X$ being the decay rate of 
$X$, which is a necessary assumption
for the generation of modulated perturbations.~\footnote{If $X$ decays very 
fast and its decay products quickly thermalize, temperature of the 
resulting (subdominant) thermal bath will have superhorizon fluctuations. 
This will lead to fluctuations in thermal corrections to the masses and 
couplings of the fields which are coupled to the thermal bath. Modulated 
perturbations 
can still be generated if the relevant process is controlled by some of these 
fields, but we do not consider this possibility here.} We also assume that the 
$X$-dependent term dominates $\Gamma_*$ and $\Gamma_0$ 
can be neglected in eq.~(\ref{rate}). For $m_X < \Gamma_*\,$,
the field $X$ is frozen at a value $X_I \pm H_I\,$ when the modulated
perturbations are generated. The assumption $X_0 < M_p$ ensures that
its energy density is subdominant at this time.~\footnote{$M_p 
= 2.4 \times 10^{18}$ GeV is the reduced Planck mass.} This will continue to
be the case provided $X$ decays sufficiently quickly. The modulated
perturbations are then estimated as $\zeta \sim H_I / X_0\,$. However, 
if $X$ survives long enough, its
energy density can overcome that of the inflaton decay products
which, as discussed in the Introduction, redshifts as radiation for
$t > t_*\,$. While $X$ is initially frozen, as soon as the Hubble
parameter drops below $m_X\,$, the field starts oscillating about
$X_0\,$. The amplitude of the oscillations decreases as $a^{-3/2}\,$,
where $a$ is the scale factor of the universe, so that the energy
density of $X$ redshifts as that of matter. The $X$-dominance occurs when the 
Hubble parameter is
\begin{figure}[tb]
\includegraphics{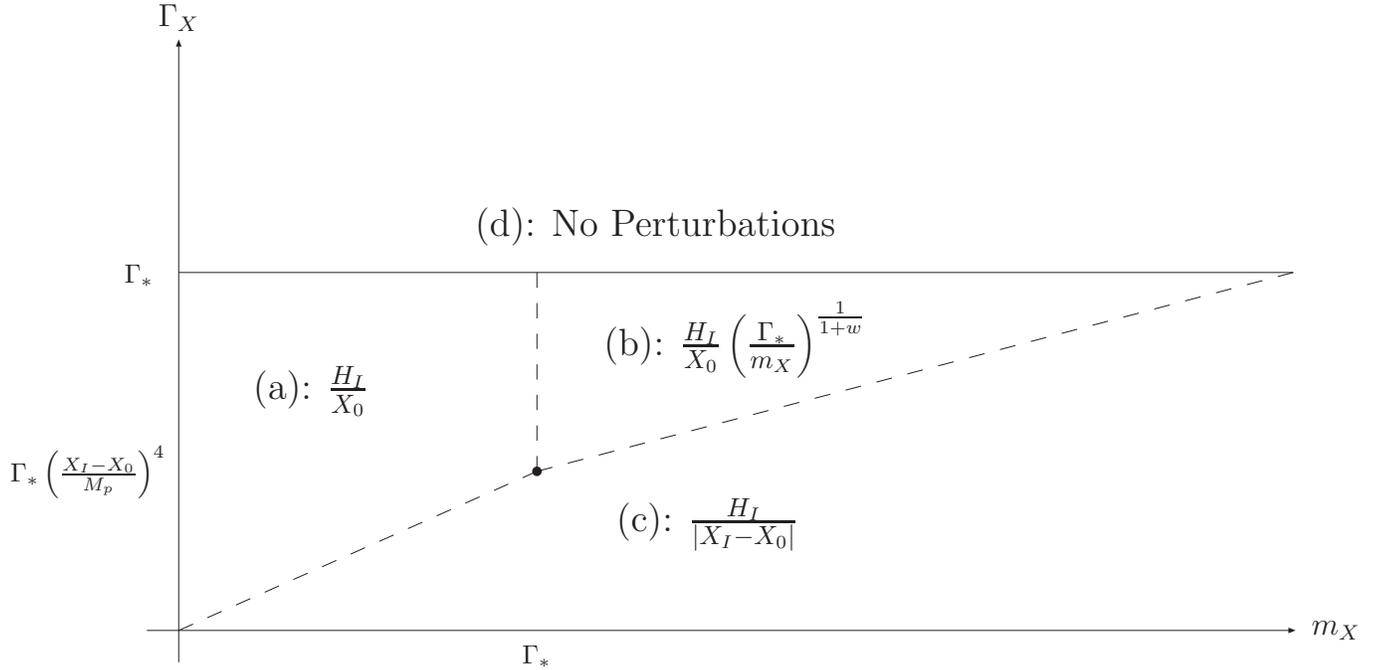}
\caption{The amplitude of perturbations when $X_0 \gg \vert X_I - X_0 
\vert\,$. In cases (a) and (b) perturbations are generated before 
and after the start of $X$ oscillations respectively. Case
(c) corresponds to $X$-dominance which is essentially 
the same as the curvaton scenario. The boundary of (c) with (a) and (b) 
is defined by $\Gamma_X = H_{eq}$, with $H_{eq}$ given in eqs.~(\ref{heq1}) 
and (\ref{heq2}) respectively.}
\label{fig2}
\end{figure}
\begin{equation} \label{heq1}
H_{\rm eq} = m_X \, \left( \frac{X_I - X_0}{M_p} \right)^4
\quad\quad\quad,\quad m_X < \Gamma_*
\end{equation}
If $\Gamma_X < H_{\rm eq}\,$, the field $X$ dominates before
decaying. As a consequence, the relevant primordial perturbations come
from fluctuations in the (potential) energy density of $X\,$;
$\rho_X \propto \left( X_I - X_0 \pm H_I \right)^2\,$, and hence
$\zeta \sim H_* / \vert X_I - X_0 \vert$ in this case.

If $m_X > \Gamma_*\,$, the analysis is complicated by the fact that
the oscillations of the $X$ field start before $t_*\,$. Dominance of $X$ now 
occurs at
\begin{equation}
H_{\rm eq} = \left( \frac{\Gamma_*}{m_X} \right)^{\frac{1-3 w}{1 + w}}
\, m_X \, \left( \frac{X_I - X_0}{M_p} \right)^4 \quad\quad\quad,\quad
m_X > \Gamma_*
\label{heq2}
\end{equation}
where $w$ is the (unknown) equation of state of the inflaton decay
products for $\Gamma_* < H < m_X\,$. If the equation of state of the universe
changes throughout this interval, the first term on the right-hand side 
of~(\ref{heq2}) will 
be replaced by the product of terms expressing the redshift in each 
sub-interval. If $\Gamma_X < H_{\rm eq}\,$, the $X$ field
dominates and perturbations in the energy density of
$X$ will give again $\zeta \sim H_I / \vert X_I - X_0 \vert
\,$. For $\Gamma_X > H_{\rm eq}\,$, however, modulated
perturbations will have a smaller amplitude than the one found
above due to the fact that $\delta X$ also decreases as
$a^{-3/2}$ during the oscillations of $X\,$.

To summarize, the amplitude of primordial perturbations will
amount to
\be
{H_I \over X_0} \left({\Gamma_* \over m_X} \right)^{1 \over 1 + w} ~~ < ~~
{H_I \over X_0} ~~ < ~~ 
{H_I \over \vert X_I - X_0 \vert},
\ee
in cases (b), (a) and (c) respectively.~\footnote{If $\Gamma_0$ dominates in 
eq.~(\ref{rate}), perturbations will be further suppressed by a factor of 
$X^n_0 
f(X_0)/\Gamma_0$ in cases (a) and (b), while remaining unchanged in case 
(c).} Since $\vert X_I - X_0 
\vert \ll X_0$, obtaining acceptable 
perturbations requires that $H_I/\vert X_I - X_0 \vert \geq 10^{-5}$. On the 
other hand, current limits on the non-gaussianity of perturbations set the 
upper bound $H_I/X_0 \lsim 10^{-1}$~\cite{non-gauss}. Note that for 
values close to this limit, perturbations of the correct size can be generated 
only when $\Gamma_* < m_X$.       

These results should be taken only as estimates of the exact value. More 
precise values can be obtained once the rate~(\ref{rate}) is specified. 
Clearly, one should not expect a discontinuity of $\zeta$. The amplitude of 
the adiabatic mode will smoothly interpolate 
between the values given in fig.~(2) and the bordering regions are 
characterized by a significant amount of isocurvature perturbations.


\subsection{Case II: $X_0 \ll \vert X_I - X_0 \vert \leq M_p$}
\begin{figure}[tb]
\includegraphics{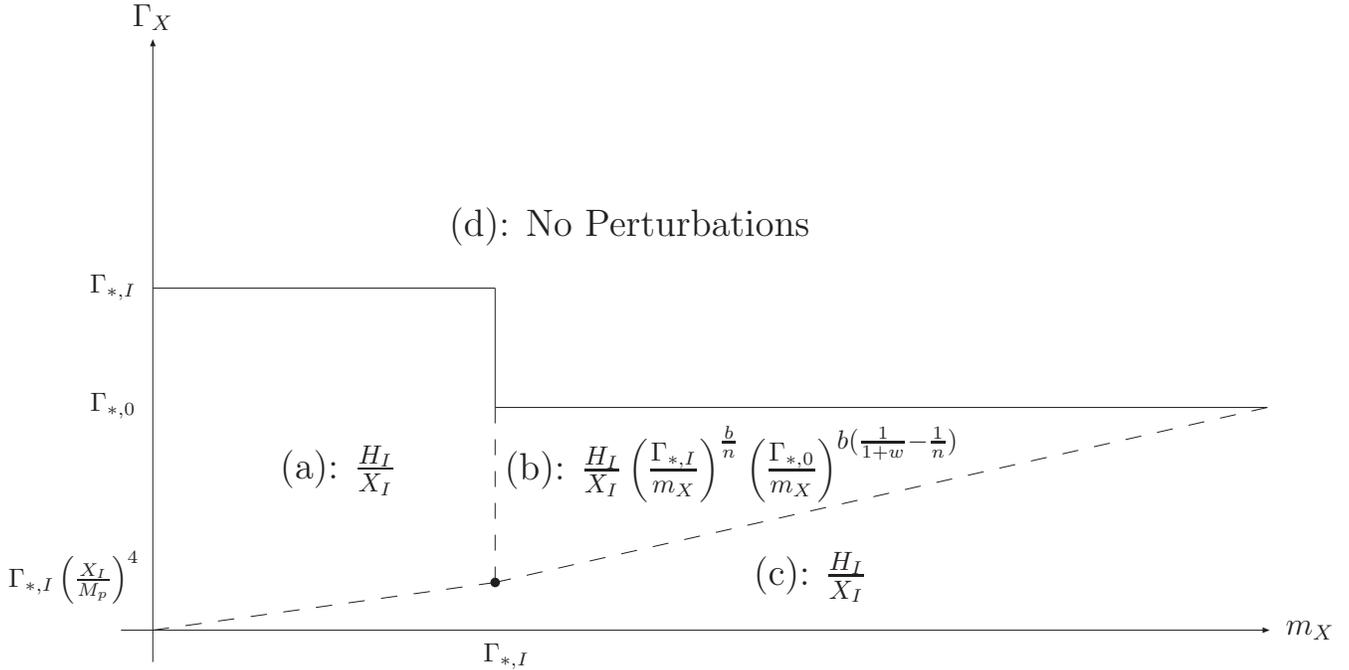}
\caption{The same as fig.~(2) but for $X_0 \ll \vert X_I - X_0 \vert\,$. 
Here $\Gamma_{*,I}$ and $\Gamma_{*,0}$ denote the initial value of $\Gamma_*$ 
and its value in the vacuum respectively. In case (b),
domination of $\Gamma_{*,0}$ by $X$-dependent and $X$-independent terms, see 
eq.~(\ref{rate}), results in $b=1$ and $b=n$ respectively.}
\label{fig3}
\end{figure}
We now discuss the case where the initial displacement of $X$ is
much greater than $X_0\,$. In addition, we focus on the situation in which
the rate~(\ref{rate}) is initially much greater than in the
vacuum of the theory, i.e. that $\Gamma_{*,I} \equiv \Gamma_* \left( X_I 
\right)
\gg \Gamma_{*,0} \equiv \Gamma_* \left( X_0 \right)\,$. Let us first
discuss the simpler case $m_X < \Gamma_{*,I}\,$,
characterized by a frozen field $X$ at $t_*\,$. In this case, the
analysis is analogous to the one in the previous subsection. If the
field $X$ never dominates, the amplitude of modulated
perturbations now amounts to $\zeta \sim H_I / X_I \,$. Remarkably,
the same value (up to an ${\cal O}(1)$ factor) is obtained also if the
field $X$ survives long enough to dominate, since $\rho \simeq \rho_X
\propto \left( X_I \pm H_I \right)^2\,$ in this case.

The situation for $m_X > \Gamma_{*,I}$ is instead more complicated. As
soon as $X$ starts oscillating with a decreasing amplitude, the
rate $\Gamma_*$ changes according to
\begin{equation}
\Gamma_{*,I} \rightarrow \Gamma \left( X \right) \propto X^n \propto
H^{\frac{n}{1+w}} \,\,,
\end{equation}
where we remind that $w$ denotes the equation of state before $t_*\,$.
If $n \geq 1+ w\,$ (expected, for example, for perturbative processes) 
$\Gamma_*$ decreases more rapidly than (or at the same rate as)
$H\,$. Hence, the process which generates modulated perturbations
will not be efficient until its rate stabilizes at
$\Gamma_{*,0}$. Note that the
situation in this case is very different from the one discussed in the previous
subsection, where $\Gamma_* \simeq \Gamma_{*,0}$ at all times. At $t =
t_*\,$, the field $X$ evaluates to
\begin{equation}
X \left( t_* \right) \simeq \left(X_I \pm H_I \right) \left(
\frac{\Gamma_{*,0}}{m_X} \right)^{\frac{1}{1+w}} \,\,.
\label{newg}
\end{equation}
First assume that $\Gamma_{*,0} \simeq X^n_0 f(X_0)$, i.e. that the 
$X$-dependent term dominates in the vacuum. Since $f(X)$ is 
assumed to be a 
slowly varying fucntion of $X$, see the discussion after eq.~(\ref{rate}), 
$X_I/X_0 \simeq (\Gamma_{*,I}/\Gamma_{*,0})^{1/n}$ when $n \geq 1+w$. Thus the 
expression for the ampltiude of perturbations can be cast in the form
\be \label{x1}
\zeta \sim {H_I \over X_I} \left({\Gamma_{*,I} \over m_X}\right)^
{1/n} \left({\Gamma_{*,0} \over m_X}\right)^{{1 \over 1+w} - {1 \over n}}.
\ee
The situation will be somewhat different when the $X$-independent term 
dominates in the vacuum. Now
$\Gamma_{*,0} \simeq \Gamma_0$, while $\delta \Gamma_{*,0} \sim X^{n-1}_I H_I
(\Gamma_{*,0}/m_X)^{n/1+w}$. This results in a smaller value for the 
amplitude of perturbations
\be \label{x2}
\zeta \sim {H_I \over X_I} \left({\Gamma_{*,I} \over m_X}\right) 
\left({\Gamma_{*,0} \over m_X}\right)^{{n \over 1+w}-1}.
\ee
If $X$ dominates, we have again $\zeta \sim H_I / X_I \,$. This occurs 
provided $X$ decays after $H_{\rm eq}\,$, given by eq.~(\ref{heq2}).
   
It turns out from (\ref{x1}) and (\ref{x2}) that $\zeta < H_I/X_I$ in case (b).
It is therefore possible to generate acceptable perturbations in this case 
for $10^{-5} < H_I/X_I \lsim 10^{-1}$. The upper bound is again given by
current limits on the non-gaussianity of perturbations. When
$H_I/X_I$ is close to this value, the modulated perturbations mechanism can 
work even if $X_I$ is 
entirely due to the accumulation of quantum fluctuations, and if 
inflation does not last much longer than $60$ e-foldings~\cite{fkl}. Note 
that the 
last two terms on the right-hand side of (\ref{x1}) and 
(\ref{x2}) should yield a number $> 10^{-4}$. Otherwise oscillations of $X$ 
will suppress modulated perturbations to an unacceptably low value.


\section{Particle physics examples of $X$} \label{examples}

In this section we present three realistic particle physics 
examples of $X$. We will discuss different possibilties for generating 
sufficient perturbations while satisfying other cosmological constraints, 
particularly to avoid the gravitino problem. As we will point out, these 
examples actually represent a wide range of particle physics candidates for 
$X$.    


\subsection{Moduli}

In superstring models physical parameters are usually set by the
VEV of moduli scalar fields, which acquire mass
only after supersymmetry is broken. In a cosmological setting,
supergravity effects typically provide these fields with a mass of the
order of the Hubble parameter $H\,$~\cite{hubble,drt}. However, there are 
situations 
in which (for example, due to a ``Heisenberg symmetry''~\cite{bg}) 
supergravity
corrections do not affect some of these directions. These fields may
then naturally be expected to have a mass of the order of the low energy
supersymmetry breaking; $m_X \sim {\cal O}(\rm TeV)$. Moduli are 
gravitationally 
coupled to the other fields in the theory, so that their decay rate
is estimated to be $\Gamma_X \sim
m^3_X/M^2_p$. They are typically characterized by $X_0 \sim M_p\,$, thus 
leading to case I in above.

Modulated perturbations can be expected if any parameter relevant for
reheating is controlled by one of these fields~\cite{kofman}. However, as we 
shall see, they turn out to be rather small in this example. The final
result for the perturbations is sensitive to the difference $\vert X_I
- X_0 \vert\,$, namely on the initial displacement of the modulus from
the minimum of the potential. We regard this quantity as a free
parameter, with the only constraint that it should be greater than the
Hubble parameter during inflation $H_I$. From eq.~(\ref{heq1}) we see
that for $\vert X_I - X_0 \vert \lsim 10^{11} \, {\rm GeV}\,$ the
modulus field decays before it dominates. However, as we just noted,
the limit on $\vert X_0 - X_I \vert$ povides an upper bound also on
$H_I$. This results in modulated perturbations that are too small, see 
fig.~(2).

For $\vert X_I - X_0 \vert \gsim 10^{11}$ the modulus decays after it
dominates, and the relevant perturbations are the ones
in the energy density of $X\,$.~\footnote{To be precise,
$m_X \gsim 50$ TeV is needed so that the moduli decay will result
in a reheat temperature $\gsim {\cal O}(\rm MeV)$, compatible with the
big bang nucleosynthesis. This can be achieved while keeping the mass
splitting between matter fields and their supersymmetric partners at
the TeV level, for example, in models of anomaly-mediated supersymmetry
breaking~\cite{anomaly}.} As long as $\vert X_I - X_0 \vert
< M_p\,$, the amplitude of the perturbations amounts to $ \sim H_I /
\vert X_I - X_0 \vert \,$ (see fig.~\ref{fig2}). Acceptable
perturbations are then generated provided $H_I \sim 10^{-5} \, \vert
X_I - X_0 \vert \,$. Actually
$m_X \gsim 50$ TeV is needed so that the moduli decay will result
in a reheat temperature $\sim {\cal O}(\rm MeV)$, compatible with the
big bang nucleosynthesis. This can be achieved while keeping the mass
splitting between matter fields and their supersymmetric partners at
TeV level, for example, in models of anomaly-mediated supersymmetry
breaking~\cite{anomaly}. For $\vert X_I - X_0 \vert > M_p\,$, the modulus
$X$ drives a stage of inflation, and the analysis of the previous
section does not apply. However, due to the smallness of $m_X$,i.e. $m_X \ll 
10^{13}$ GeV, 
identification of modulus with the inflaton field results in unacceptably 
small perturbations.


\subsection{Right-handed sneutrinos}

The right-handed (RH) sneutrinos arise in supersymmetric
extensions of the standard model~\cite{nilles} which explain the
smallness of the mass of the left-handed (LH) neutrinos via the
see-saw mechanism~\cite{see-saw}. A non-zero VEV for ${\tilde N}$ at the 
minimum of its potential breaks $R$-parity, thus destabilizing the
lightest supersymmetric particle (LSP)~\cite{nilles}. This VEV should
be $\ll 1$ GeV, if the LSP survives until today and
constitutes the dark matter. Nevertheless, since $\tilde N$ is a
standard model gauge singlet, it is conceivable that the $X$-indepdendent
contribution to the process which generates modulated perturbations
can be recasted in a much larger $X_0\,$, as we have discussed after
eq.~(\ref{model}). Consider an example where
perturbations are imprinted at the inflaton decay which proceeds
through the superpotential couplings (the boldface charachters denote
superfields)
\be \label{example}
{{\bf X} \over M_p} \Phi \Psi \Psi + h \Phi \Psi \Psi.
\ee
Then $X_0 = h M_p$, leading to case I (II) in above if $X_I \ll h M_p$ ($\gg h
M_p$). We concentrate on the latter possibility as the former will lead to a 
situation similar to that discussed in the previous subsection.~\footnote{
Generating density perturbations from the RH sneutrinos has also been 
considered in~\cite{anupam}.}

The see-saw formula gives~\cite{see-saw}
\begin{equation}
m_X \geq y^2 \langle H \rangle^2 / m_{\nu} \sim y^2 \, 10^{15} \, {\rm GeV}
\,\,,
\label{see-saw}
\end{equation}
where $\langle H \rangle \sim 200$ GeV is the electroweak symmetry breaking 
scale, $m_\nu \simeq 0.05$ eV is the mass of the heaviest light neutrino and 
$y$ denotes the neutrino Yukawa coupling. The same interactions give rise to 
a sneutrino decay rate
\begin{equation} \label{yukawa}
\Gamma_X = \frac{y^2}{8 \, \pi} m_X \sim y^4 \, 10^{14} \, {\rm GeV}
\,\,.
\end{equation}
In realistic models of neutrino masses based on the see-saw 
mechansim~\cite{neutrino}, $m_X$ is typically 
much larger than the scale of 
electroweak symmetry breaking. Moreover, producing sufficient baryon 
asymmetry via leptogenesis~\cite{fy} leads to additional constraints on the 
model parametrs. In particular, various
leptogenesis scenarios~\cite{plumacher,sacha,giudice,hmy,ad3} require that 
$m_X > 10^5$ GeV (for thermal leptogenesis the bound is about $4$ order of 
magnitude stronger) unless some specific fine-tuning occurs (for example, 
having nearly degenerate RH (s)neutrinos~\cite{degenerate}). Therefore, $m_X 
\gg 1$ TeV in general.

In principle, $\Gamma_*$ can be larger or smaller than $m_X\,$. In the
former case, modulated perturbations must be generated by 
a very rapid process, most notably non-perturbative inflaton 
decay via preheating. A large $\Gamma_*\,$ may be associated with a high 
reheat temperature $T_R\,$,~\footnote{We have defined $t_*$ as the moment
when the inflaton decay products have an equation of state of
radiation. This does not necessarily mean that their distribution is
thermal. Once thermalization of decay products has completed, $T_R$ will be 
the highest temperature in the radiaiton-dominated era.} which can then lead 
to thermal
overproduction of gravitinos. In gravity-mediated models of
supersymmetry breaking the gravitino mass is $m_{3/2} \simeq 100~{\rm
GeV}-1$ TeV and (up to logarithmic factors)~\cite{ellis}
\be \label{gravitino1}
{n_{3/2} \over s} \simeq 10^{-12} \left({T_R \over 10^{9}~{\rm
GeV}}\right).
\ee
Bounds from nucleosynthesis (due to photodissociation of light
elements by gravitino decay products) impose $T_R \leq 10^9$ GeV~\cite{cefo},
corresponding to $H_R \leq 10$ GeV ($H_R$ being the Hubble parameter
at $T = T_R$). In models of gauge-mediated supersymmetry breaking,
the gravitino is the LSP and $m_{3/2}$ can be as low as $1$ KeV. Its
fractional energy density is in this case~\cite{mmy,dmm}
\be \label{gravitino2}
\Omega_{3/2} h^2 \simeq 0.8 \left({M_3 \over 1~{\rm TeV}}\right)^2 
\left({10~{\rm MeV} \over m_{3/2}}\right) \left({T_R \over 
10^{6}~{\rm GeV}}\right),
\ee
where $h \approx 0.7$ and $M_3$ is the gluino mass parameter. Then,
for a typical value $M_3 \sim 500$ GeV, the dark matter limit results in 
$T_R \lsim
10^{8} \,m_{3/2}$, corresponding to $H_R \leq 10^{17} m^2_{3/2}/M_p$. In both 
cases higher $T_R$ can be accommodated, provided the gravitino abundance
is diluted by an entropy release at later times. This entropy can be
naturally provided by the decay of the sneutrino if its
energy density becomes significantly greater than that of the thermal 
bath. In addition, the observed baryon 
asymmetry of the universe can also be generated by such a late 
decay~\cite{hmy,ad3}. Then~(\ref{heq1}), (\ref{yukawa}), 
(\ref{gravitino1}) and (\ref{gravitino2}) lead to the bounds
\be \label{decgr}
y \leq 5 \left({10^9~{\rm GeV} \over T_R}\right) \left({X_I - X_0 
\over M_p}\right)^2,
\ee
and 
\be \label{decga}
y \leq 5 \times 10^8 \left({m_{3/2} \over T_R}\right) \left({X_I - X_0 
\over M_p}\right)^2
\ee
in gravity-mediated and gauge-mediated models respectively. We are thus 
naturally led to consider the case in which the sneutrino dominates. 
Primordial perturbations will then be related to fluctuations in the energy 
density of $X\,$, corresponding to case (c) in fig.~(3).

Let us finally consider the case where $\Gamma_{*,I} < m_X\,$.
If the reheat temperature respects the gravitino bound, no entropy generation 
will be required. Then the sneutrino may decay before or after dominating the 
universe corresponding to cases (b) and (c) in fig.~(3), respectively, 
and perturbations will be given by~(\ref{x1}) in the former case. If 
the reheat temperature exceeds the gravitino bound, sneutrino 
dominance will be favoured again to solve the gravitino problem.


\subsection{Supersymmetric flat directions}

There are many directions in the 
space of the Higgs, slepton and squark fields along which the scalar 
potential of the MSSM identically 
vanishes in the limit of exact supersymmetry~\cite{flat}. These flat 
directions 
acquire a mass from supersymmetry breaking. In gravity-mediated models 
$m_X = 100~{\rm GeV}-1$ TeV~\cite{nilles}. In gauge-mediated models $m_X 
\simeq 100~{\rm GeV}-1$ TeV for small $\langle X \rangle$, it drops 
$\propto \langle X \rangle^{-1}$ at intermediate VEVs and $m_X 
\sim m_{3/2}$ for large $\langle X \rangle$~\cite{dmm}. 

The Higgs fields have a VEV 
$\sim 100$ GeV in the vacuum, while those of the sleptons and squarks vansih. 
Therefore identification of $X$ with the MSSM flat directions (the primary 
example considered in~\cite{dgz}) leads to case II in above where 
perturbations read from fig.~(3). The situation is qualitatively similar to 
the sneutrino example but some differences exist. For the flat directions 
$m_X$ is typically much smaller 
than the RH sneutrinos, particularly in gauge-mediated models. This allows 
a smaller $\Gamma_*$ (as expected for perturbative processes)
when $\Gamma_{*,I} > m_X$. The fact that the Higgs, slepton and 
squark fields are gauge non-singlets can also result in a different situation 
when $\Gamma_{*,I} < m_X$. Consider again the example of 
inflaton decay in eq.~(\ref{example}), with $X$ being a Higgs, slepton 
or squark field. Now the two terms in~(\ref{example}) 
must couple the inflaton to different final states $\Psi_1$ and $\Psi_2$, 
resulting in decay rates $\Gamma_X$ and $\Gamma_0$ respectively. 
In any acceptable scenario $\Gamma_{*,0} \simeq \Gamma_0$, and hence 
perturbations will be given by~(\ref{x2}), instead of~(\ref{x1}) for the 
sneutrinos, provided $X$ does not dominate. 

The $X$-dominance leads to case (c) in fig.~(3). The decay of the flat 
directions proceeds through the SM Yukawa and gauge couplings. In particular, 
one may wonder whether $X$ can dominate at all as gauge couplings lead to a 
rapid decay of $X$. However, the gauge (and Yukawa) couplings also result in 
an effective mass $m_{eff}$ for the decay prodcuts due to their couplings to 
the $X$ condensate and thermal effects. The decay is 
kinematically forbidden so long as $m_{eff} > m_X$~\cite{linde}. It 
occurs when $H = H_d \equiv {\rm min} [\Gamma_X,H_{kin}]$, with 
$H_{kin}$ denoting the value of Hubble parameter at which $m_{eff}$ drops 
below $m_X$.~\footnote{For many 
flat directions oscillations of the zero-mode condensate can be fragmented 
into non-topological solitons, so called supersymmetric Q-balls~\cite{q}, 
which may be long-lived or stable. This will
further complicate the situation, and hence we do not 
consider it here.} Therefore $X$-dominance actually requires that $H_d < 
H_{eq}$, where $H_{eq}$ is given by~(\ref{heq2}).


\subsection{Summary}

In supergravity models all scalar fields typically 
receive a supersymmetry breaking mass $\sim m_{3/2}$. If a supersymmetry 
conserving (superpotential) mass term is 
allowed for a field, its mass can be $\gg m_{3/2}$. However, a tree-level 
(supersymmetry breaking or conserving) mass term will 
be forbidden if the theory has some symmetry in the superpotential (thus 
resulting in a Goldstone boson) or in the kinetic function (as in no-scale 
supergravity~\cite{no-scale}). In this case the corresponding field 
obtains a mass $\ll m_{3/2}$ through symmetry breaking interactions at higher 
orders. Thus, unless forbidden by some symmetry, $m_X \geq m_{3/2}$ will 
be expected in general. Under the most 
general circumstances, a massive scalar field is either in a hidden 
sector, thus gravitationally coupled to the SM fields, or has 
gauge and/or Yukawa couplings to the observable sector. Moduli are an example 
of the case with $m_X \sim m_{3/2}$ and gravitational decay to matter fields. 
The supersymmetric flat directions ($m_X \geq m_{3/2}$) and the RH sneutrinos 
($m_X \gg m+{3/2}$) have gauge and Yukawa 
couplings to the SM fields. The three 
examples considered above therefore represent a wide range of possible $X$ 
candidates.        

One comment is in order before closing this section. We have 
only considered the mass term in $V(X)$, see~(\ref{model}). However, higher 
order terms are naturally expected to arise and make the potential steeper 
than $X^2$ at large field values. This will result in an upper limit 
$X_{max}$ on $X_I$ at which $V''(X_{max}) \sim H^2_I$. When $X_I \simeq 
X_{max}$, fluctuations of $X$ are attenuated 
even before oscillations start~\cite{emp}. This will suppress the amplitude of 
perturbations compared to that is given in figs.~(2) and (3), which 
will be valid so long as $X_I < X_{max}$. A limit on 
$X_I$, through the expressions for the amplitude of perturbations, translates 
into an upper bound on the scale of inflation $H_I$. The strongest bound $H_I 
< 10^{-5} X_{max}$ is obtained when $\zeta \sim H_I/X_I$, while cases with 
$\zeta < H_I/X_I$ lead to weaker constraints on $H_I$. In particular, $H_I$ 
can be just one order of magnitude below $X_{max}$ when the value of 
$H_I/X_I$ is close to 
the current limit from non-gaussianity of perturbations. The 
bound on $H_I$ can in this case be comparable with (or weaker than) $H_I < 
10^{13}$ GeV, which is required so that the inflaton fluctuations not 
yield too large perturbations, even if $X_{max} \ll M_p$. The value of 
$X_{max}$, 
signifying the importance of higher order terms, strongly depends on the 
particle physics identification of $X$. For a modulus field these terms are 
$M_p$ suppressed and the mass term will be 
dominant up to $X_I \lsim M_p$ . For the RH sneutrinos and supersymmetric flat 
directions, $X_{max}$ can assume a wide range of values $\ll M_p$. Due to 
the model-dependence of $X_{max}$ and the constraint it imposes on $H_I$, we 
have not considered higher order terms in $V(X)$ in our discussion.


\section{Discussion} \label{discuss}
   
So far we have assumed that only the inflaton and the $X$ field 
dynamically evolve in the early universe. Any other scalar fields with masses 
smaller than $H_I$ are also expected to acquire inflationary fluctuations, and 
be substantially displaced from the minimum of their potential at early times. 
They can then contribute to density perturbations in a similar fashion as the 
$X$ field. Since $X$ is considered to be the dominant source for 
generating perturbations here, fluctuations of other scalar fields should be 
suppressed. This is particularly true for $X$ decay products, generally 
denoted by $\chi$. This can be naturally 
achieved if these fields have a mass $\gsim H_I$ during inflation. One 
possibility is that supergravity effects provide such a 
mass~\cite{hubble,drt}. Note, however, that the supergravity mass for 
$X$ should be $< H_I$. As pointed out earlier, this can happen if $X$ has a 
different kinetic function from other fields. It is also possible that 
different masses for different fields arise dynamically as a result of quantum 
corrections~\cite{gmo,adm}. Fluctuations of $\chi$ will also be suppressed if 
its coupling to $X$ and/or inflaton, denoted by $y$ and $h$ respectively, is 
sufficiently large, i.e. that $y X_I > H_I$ and $h \phi_0 > H_I$ respectively 
($\phi_0$ is the initial amplitude of inflaton oscillations). Since $m_X < 
H_I$, preheating decay of $X$ and/or inflaton to $\chi$ will be expected in 
this case~\cite{preheat}. 

The discussion of Section~\ref{examples}, together with
the requirement for suppressing fluctuations of $\chi$, lead us to the 
following ``benchmark scenarios'' for modulated perturbations:             
\begin{enumerate}
\item
$X$ is in the hidden sector and $m_X \gsim 50$ TeV. Supergravity 
effects provide a mass $> H$ for $\chi$ suppressing its fluctuations. $X$ 
dominates the universe and fluctuations in its energy density give 
rise to perturbations; case (c) in fig.~(2). The situation is essentially the 
same as the curvaton scenario.
\item
$X$ is in the observable sector and $m_X \gg 1$ TeV. Perturbations are 
generated when the inflaton decyas via preheating, before $X$ starts 
oscillating; case (a) in fig.~(3). 
Coupling of 
$X$ to $\chi$ is large enough ($y > 10^{-5}$) to suppress its fluctuations. 
This also leads to a rapid non-perturbative decay of $X$ to $\chi$, which is 
welcome as $X$ will not dominate and consequently affect perturbations. The 
universe thermalizes sufficiently late, thus there will be no thermal 
overproduction of gravitinos and no late time entropy generation 
will be required.     
\item
$X$ is in the observable sector and starts oscillating 
before perturbations are imprinted; case (b) in fig.~(3). This happens due to 
the slowness of the process generating perturbations, for example, 
gravitational decay of the inflaton or decay of a long-lived massive particle 
produced in the inflaton decay. Such a slow process will be required by the 
gravitino bound when the universe thermalizes rapidly. Oscillations of $X$ in
this case suppress modulated perturbations compared to that in the previous 
scenario. Preheating decay 
of $X$ is not allowed, thus $y X_I < m_X$, since it should 
survive long enough until perturbations are generated. 
This rules out 
suppressing $\chi$ fluctuations through its coupling to $X$ (supergravity 
effects or a large coupling to the inflaton can do this instead). On the other 
hand, since fluctuations in the energy density of $X$ can be $\gg 10^{-5}$,
$X$ must decay long before dominating in order not to affect 
perturbations. This, after using~(\ref{heq2}), will result in an upper bound 
on $y$. 

We note that $X$ can undergo early oscillating if it acquires a large 
thermal~\cite{thermal} or non-thermal~\cite{nonthermal} mass $m_{X,eff}$. 
This will happen provided 
$\chi$ are in equilibrium with the (p)reheat plasma and $y$ is sufficiently 
large,~\footnote{$y$ should not be too large, however, otherwise 
re-scatterings by $\chi$ quanta will quickly destroy the $X$ condensate.} so 
that $m_{X,eff}$ exceeds the Hubble parameter when $H > m_X$.        
\item
$X$ is in the observable sector and starts oscillating before or after 
modulated perurbations are generated. The reheat temperature exceeds the 
gravitino bound but $X$ dominates the universe before decaying, thus diluting 
the gravitinos in excess. Fluctuations of $\chi$ will in this case be 
suppressed by supergravity effects or a large coupling to the inflaton. 
Perturbations will be due to fluctuations in the energy density of $X$; case 
(c) in fig.~(3).
\end{enumerate}
%
                          
\section{Summary and conclusions} \label{end}

In this paper we have studied different scenarios of modulated perturbations. 
In this mechanism adiabatic cosmological perturbations are 
generated during reheating when the equation of 
state in the different parts of the universe changes at different moments. 
The inhomogeneity arises due to inflationary fluctuations of some 
scalar field $X$ whose VEV controls the value of 
parameter(s) (such as mass or coupling) involved in the relevant process. 
This is a generic property expected in models based on superstring and 
supersymmetric 
theories. Depending on the values of the rate of the process 
$\Gamma_*$, mass of the scalar field $m_X$ and its decay rate $\Gamma_X$, 
different situations can arise. We presented a general analysis of possible 
scenarios, with the main results summarized in fig.~(2) and fig.~(3). We also 
introduced some specific particle physics examples including the string 
moduli, RH sneutrinos and supersymmetric flat directions.  
These examples, in which $m_X$ is of the same order or greater 
than the gravitino mass $m_{3/2}$, represent a wide range of particle 
physics candidates of $X$.         

The simplest possibility is that perturbations are generated when $X$ is 
frozen at an initial value $X_I$. If $m_X \geq {\cal O}(\rm TeV)$ (as 
for the RH sneutrinos and supersymmetric flat directions), this 
requires a very rapid process, namely non-perturbative inflaton decay 
via preheating. Moreover, $X$ must decay before dominating the energy 
density of the universe, and the reheat temperature should be sufficiently 
low to avoid overproduction of gravitinos. For a slower (likely perturbative) 
process, $X$ can start oscillating around the minimum of its potential before 
perturbations are imprinted. Then the expansion of the universe redshifts 
fluctuations of $X$ and dampens perturbations. An interesting consequence is 
that the level 
of non-gaussianity can in this case be just below the current limits and 
accessible to future experiments. If $X$ is long-lived (for 
example, a modulus field), it can dominate the universe 
before decaying. In this case the situation will essentially be the same as 
the curvaton scenario and fluctuations in the energy density of $X$ give rise 
to perturbations. The $X$-dominance can be a virtue when the reheat 
temperature exceeds the gravitino bound necessitating late time entropy 
production. The amplitude of the adiabatic mode is in general very different 
from one case to another. It interpolates between the different values and 
bordering regions are charachterized by a significant isocurvature component.  

In conclusion, the modulated perturbations mechanism is a viable alternative 
with potentially interesting observational consequences. It can be 
implemented in various ways and a successful particle 
physics identification can lead us to a clearer and more complete picture of 
reheating.

\section{Acknowledgements}

The author wishes to thank L. Kofman and M. Peloso for many valuable 
discussions and collaboration at earlier stages of this work, and R. Cyburt 
for careful reading of the manuscript. He also 
acknowledges the kind hospitality by CITA while this work was being 
completed. The research of R.A. is supported by the National Sciences and 
Engineering Research Council of Canada.



\begin{references}
%
\bibitem{inflation} 
For reviews on inflation, see: A. D. Linde, {\it Particle Physics and
Inflationary Cosmology}, Harwood (1990);\\
D. H. Lyth and A. Riotto, Phys. Rep. {\bf 314}, 1 (1999).
%
\bibitem{wmap}
C. L. Bennett, {\it et al}, Ap. J. S. {\bf 148}, 1 (2003); D. N. Spergel, 
{\it et al}, Ap. J. S. {\bf 148}, 175 (2003).
%
\bibitem{reheat}
A. Dolgov and A. D. Linde, Phys. Lett. B {\bf 116}, 329 (1982); L. F. Abbott, 
E. Farhi and M. Wise, Phys. Lett. B {\bf 117}, 29 (1982).   
%
\bibitem{preheat} 
L. Kofman, A. D. Linde and A. A. Starobinsky,
Phys. Rev. Lett. {\bf 73}, 3195 (1994), and Phys. Rev. D {\bf 56}, 3258 (1997).
%
\bibitem{non}
G. Lazarides and Q. Shafi, Phys. Lett. B {\bf 258}, 305 (1991); E. W. Kolb, 
A. D. Linde and A. Riotto, Phys. Rev. Lett. {\bf 77}, 4290 (1996). 
%
\bibitem{dgz}
G. Dvali, A. Gruzinov and M. Zaldarriaga, Phys. Rev. D {\bf 69}, 023505, 
(2004); {\it ibid} {\bf 69}, 083505 (2004).
%
\bibitem{kofman}
L. Kofman, astro-ph/0303614.
%
\bibitem{postma}
M. Postma, J. Cosmol. Astropart. Phys. {\bf 0403}, 006 (2004).
%
\bibitem{gruzinov}
A. Gruzinov, astro-ph/0401407.
%
\bibitem{curvaton}
D. H. Lyth and D. Wands, Phys. Lett. B {\bf 524}, 5 (2002); D. H. Lyth, C. 
Ungarelli and D. Wands, Phys. Rev. D {\bf 67}, 023503 (2003). 
%
\bibitem{moroi}
T. Moroi and T. Takahashi, Phys. Lett. B {\bf 522}, 215 (2001), Erratum-ibid 
B {\bf 539}, 303 (2002); Phys. Rev. D {\bf 66}, 063501 (2002).
%
\bibitem{nicola}
N. Bartolo and A. R. Liddle, Phys. Rev. D {\bf 65}, 121301 (2002).
%
\bibitem{dim}
K. Dimopoulos, G. Lazarides, D. H. Lyth and R. Ruiz de Austri, Phys. Rev. D 
{\bf 68}, 123515 (2003). 
%
\bibitem{gk}
P. Greene and L. Kofman, Phys. Lett. B {\bf 448}, 6 (1999), and Phys. Rev. D 
{\bf 62}, 123516 (2000). 
%
\bibitem{peloso}
M. Peloso and L. Sorbo, J. High Energy Phys {\bf 0005}, 016 (2000). 
%
\bibitem{non-gauss}
E. Komatsu, D. N. Spergel and B. D. Wandelt, astro-ph/0305189.
%
\bibitem{fkl}
G. Felder, L. Kofman and A. D. Linde, J. High Energy Phys. {\bf 0002}, 027 
(2000).
%
\bibitem{hubble}
M. Dine, W. Fischler and D. Nemechansky,
Phys. Lett. B {\bf 138}, 169 (1984); S. Bertolami and G. G. Ross,
Phys. Lett. B {\bf 183}, 163 (1987); G. Dvali, Phys. Lett. B {\bf 355}, 
78 (1995).
%
\bibitem{drt}
M. Dine, L. Randall and S. Thomas, Phys. Rev. Lett. {\bf 75}, 398 (1995), and 
Nucl. Phys. B {\bf 458}, 291 (1996).
%
\bibitem{bg}
P. Binetruy and M. K. Gaillard, Phys. lett. B {\bf 195}, 382 (1987).
%
\bibitem{anomaly}
G. F. Giudice, M. A. Luty, H. Murayama and R. Rattazzi, J. High Energy Phys. 
{\bf 9812}, 027 (1998).
%
\bibitem{nilles}
For a review on supersymmetry, see: H. P. Nilles, Phys. Rept. 
{\bf 110}, 1 (1984).
%
\bibitem{see-saw}
M. Gell-Mann, P. Ramond and R. Slansky, in {\it Supergravity},
eds. P. van Nieuwenhuizen and D. Z. Freedman (North Holland 1979);
T. Yanagida, Proceedings of {\it Workshop on
Unified Theory and Baryon number in the Universe}, eds.
O. Sawada and A. Sugamoto (KEK 1979);
R. N. Mohapatra and G. Senjanovic, Phys. Rev. Lett. {\bf 44}, 912
(1980).
%
\bibitem{neutrino}
For a review on neutrino mass models, see: S. F. King, Rept. Prog. Phys. 
{\bf 67}, 107 (2004), and references therein. 
%
\bibitem{fy}
M. Fukugita and T. Yanagida, Phys. Lett. B {\bf 174}, 45 (1986). 
%
\bibitem{plumacher}
W. B\"uchmuller, P. Di bari and M. Pl\"umacher, Nucl. Phys. B {\bf 643}, 367 
(2002); Nucl. Phys. {\bf 665}, 445 (2003), and hep-ph/0401240.
%
\bibitem{sacha}
S. Davidson, J. High Energy Phys. {\bf 0303}, 037 (2003). 
%
\bibitem{giudice}
G. F. Giudice, A. Notari, M. Raidal, A. Riotto and A. Strumia, Nucl. Phys. B 
{\bf 685}, 89 (2004). 
%
\bibitem{hmy}
K. Hamaguchi, H. Murayama and T. Yanagida, Phys. Rev. D {\bf 65}, 043512 
(2002).
%
\bibitem{ad3}
R. Allahverdi and M. Drees, Phys. Rev. D {\bf 69}, 103522 (2004).
%
\bibitem{degenerate}
A. Pilaftsis, Phys. Rev. D {\bf 56}, 5431 (1997), and Int. J. Mod. Phys. A 
{\bf 14}, 1811 (1999); A. Pilaftsis and T. E. J. Underwood, hep-ph/0309342. 
%
\bibitem{anupam}
A. Mazumdar, Phys. Lett. B {\bf 580}, 7 (2004), and Phys. Rev. Lett. 
{\bf 92}, 241301 (2004). 
%
\bibitem{ellis}
M. Yu. Khlopov and A. D. Linde, Phys. Lett. B {\bf 138}, 265 (1984); 
J. Ellis, J. E. Kim and D. V. Nanopoulos, Phys. Lett. B {\bf 145},
181 (1984);
J. Ellis, D. V. Nanopoulos, K. A. Olive and S- J. Rey, Astropart. Phys.
{\bf 4}, 371 (1996);\\
for a recent calculation, see: M. Boltz, A. Brandenburg and
W. B\"uchmuller, Nucl. Phys. B {\bf 606}, 518 (2001).
%
\bibitem{cefo}
R. H. Cyburt, J. Ellis, B. D. Fields and K. A. Olive, Phys. Rev. D {\bf 67}, 
103521 (2003).\\
for a review, see: S. Sarkar, Rept. Prog. Phys. {\bf 59}, 1493 (1996). 
%
\bibitem{mmy}
T. Moroi, H. Murayama and M. Yamaguchi, Phys. Lett. B {\bf 303}, 289 (1993).
%
\bibitem{dmm}
A. de Gouv\^ea, T. Moroi and H. Murayama, Phys. Rev. D {\bf 56}, 1281 (1997).
%
\bibitem{flat}
T. Gherghetta, C. Kolda and S. Martin, Nucl. Phys. B {\bf 468}, 37 (1996): \\
for a review on cosmological consequences of flat directions, see: K. Enqvist 
and A. Mazumdar, Phys. Rep. {\bf 380}, 99 (2003).    
%
\bibitem{q} 
A. Kusenko and M. Shaposhnikov, Phys. Lett. B {\bf 418}
(1998) 46; K. Enqvist and J. McDonald, Phys. Lett. B {\bf 425} (1998)
309. 
%
\bibitem{linde}
A. D. Linde, Phys. Lett. B {\bf 160}, 243 (1985). 
%
\bibitem{no-scale}
E. Cremmer, S. Ferrara, C. Kounnas and D. V. Nanopoulos, Phys. Lett. B {\bf 
133}, 61 (1983).  
%
\bibitem{emp}
K. Enqvist, A. Mazumdar and M. Postma, Phys. Rev. D {\bf 67}, 121303 (2003).
%
\bibitem{gmo}
M. K. Gaillard, H. Murayama and K. A. Olive, Phys. Lett. B {\bf 355}, 71 
(1995).
%
\bibitem{adm}
R. Allahverdi, M. Drees and A. Mazumdar, Phys. Rev. D {\bf 65}, 065010 (2002).
%
\bibitem{thermal}
R. Allahverdi, B. A. Campbell and J. Ellis, Nucl. Phys. B {\bf 579}, 355 
(2000); A. Anisimov and M. Dine, Nucl. Phys. B {\bf 619}, 729 (2001).
%
\bibitem{nonthermal}
L. Kofman, A. D. Linde and A. A. Starobinsky, Phys. Rev. Lett. {\bf 76}, 1011 
(1996); I. I. Tkachev, Phys. Lett. B {\bf 376}, 35 (1996). 
%

\end{references}
\end{document}